\documentclass[preprint,aps,eqsecnum,superscriptaddress]{revtex4}
\usepackage{graphicx,amsfonts}
\usepackage{amsmath}
\usepackage{slashed}

\newcommand{\be}{\begin{equation}}
\newcommand{\ee}{\end{equation}}
\newcommand{\bea}{\begin{eqnarray}}
\newcommand{\eea}{\end{eqnarray}}
\newcommand{\bml}{\begin{subequations}}
\newcommand{\eml}{\end{subequations}}
\newcommand{\pa}{\partial}

\newcommand{\vx}{\vec{x}}

\newcommand{\vk}{\vec{k}}

\newcommand{\ve}{\varepsilon}

\begin{document}

\title {Born series and unitarity in noncommutative quantum mechanics}

\author{F. S. Bemfica}
\author{H. O. Girotti}
\affiliation{Instituto de F\'{\i}sica, Universidade Federal do Rio Grande
do Sul, Caixa Postal 15051, 91501-970 - Porto Alegre, RS, Brazil}
\email{fbemfica, hgirotti@if.ufrgs.br}

\begin{abstract}

This paper is dedicated to present model independent results for noncommutative quantum mechanics. We determine sufficient conditions for the convergence of the Born series and, in the sequel, unitarity is proved in full generality.
\end{abstract}

\maketitle
\newpage

\section{Introduction}
\label{sec:level1}

In this work we shall be concerned with quantum systems whose dynamics is described by a self-adjoint Hamiltonian $H(Q,P)$ made up of the Cartesian coordinates $Q^l, l= 1,\ldots, N$ and their canonically conjugate momenta $P^j, j = 1,\ldots, N$. However, unlike the usual case, coordinates and momenta are supposed to obey the non-canonical equal-time commutation rules

\bml
\label{1}
\bea
&&\left[Q^l, Q^j\right] = -2 i\hbar\theta^{lj},\label{mlett:a1}\\
&&\left[Q^l, P_j\right] = i\,\hbar\,\delta^{l}_{\,\,j},\label{mlett:b1}\\
&&\left[P_l, P_j\right] = 0.\label{mlett:c1}
\eea
\eml

\noindent
The distinctive feature is, of course, that the coordinate operators do not commute among themselves. The lack of non-commutativity of the coordinates is parameterized by the real antisymmetric $N \times N$ constant matrix $\|\theta\|$\cite{footnote1}. In Refs.\cite{Chaichian1,Gamboa1,Gamboa2,Girotti1,Bemfica} one finds specific examples of noncommutative systems whose quantization has been carried out\cite{footnote2}. While Ref.\cite{Chaichian1} is concerned with the distortion provoked by the non-commutativity on the spectrum of the hydrogen atom, Refs.\cite{Gamboa1,Gamboa2,Girotti1} deal with the noncommutative two-dimensional harmonic oscillator, an exactly solvable model. In Ref.\cite{Bemfica} the authors elaborate about the effects of the non-commutativity in the case of a multi-particle system: the electron gas.

However, model independent developments concerned with the backbones of quantum mechanics are still lacking. Our purpose in this work is to contribute to fill up this gap.

In Section 2 we briefly quote, for future purposes, the main steps leading to the formulation of the quantum dynamics of noncommutative systems in the Schrödinger picture.

Sections 3 and 4 contain the original results in this work. In section 3 we determine sufficient conditions for the existence of the Born series, while in Section 4 we prove that perturbative unitarity holds.

The conclusions are contained in Section 5.

\section{Quantum dynamics of noncommutative systems in the Schr\"odinger picture}
\label{sec:level2}

A representation of the algebra in Eq.(\ref{1}) can be obtained by writing

\bml
\label{II-11}
\bea
&& Q^l\,\equiv\,X^l\,+\,\theta^{lj}\,K_j\,,\label{mlett:aII-11}\\
&& P_l\,\equiv\,K_l\,,\label{mlett:bII-11}
\eea
\eml

\noindent
where the $X$'s and $K$'s obey the canonical commutation relations

\bml
\label{II-12}
\begin{eqnarray}
&&\left[X^{l}, X^{j}\right]\, = \,0\,,\label{mlett:aII-12}\\
&&\left[X^{l}, K_{j}\right] = i\,\hbar\,\delta^{l}_{\,\,j}\,,\label{mlett:bII-12}\\
&&\left[K_{l}, K_{j}\right] = 0\,.\label{mlett:cII-12}
\end{eqnarray}
\eml

\noindent
Clearly, the common $X$-eigenvectors ($|{\vec x}> \equiv | x^{1},\ldots,x^{l},\ldots,x^{N}>$) provide a basis in the space of states for representing the algebra (\ref{1}). Thus, for a Hamiltonian

\be
\label{II-14}
H(Q ,P) = \frac{P_l P_l}{2 M} + V(Q)
\ee

\noindent
and, therefore,

\be
\label{II-15}
H(X^{l}\, + \,\theta^{lj}\, K_{j}\,,\, K_{l}) = \frac{K_l K_l}{2 M} + V(X^{l} + \theta^{lk}\, K_k)\,,
\ee

\noindent
it has been shown elsewhere\cite{Gamboa1,Gamboa2,Girotti1} that the time evolution of the system, in the Schr\"odinger picture, is described by the wave equation

\be
\label{II-16}
- \frac{\hbar^2}{2M} \nabla_{x}^2 \Psi(x,
t) + V (x)\star \Psi(x, t) = i \hbar \frac{\pa
 \Psi(x, t)}{\pa t}\,,
\ee

\noindent
where $\nabla_{x}^2$ designates the $N$th-dimensional Laplacian, $M$ is a constant with dimensions of mass while $\star$ denotes the Gr\"onewold-Moyal product\cite{Gronewold,Moyal,Filk1}, namely,

\bea
\label{II-17}
V (x)\star \Psi(x, t)&\equiv& V(x) \left[\exp \left(- i
\hbar \overleftarrow{\frac{\pa}{\pa
x^l}} \,\theta^{lj}\, \overrightarrow{\frac{\pa}{\pa
x^{j}}}\right)\right] \Psi(x, t)\nonumber\\
&=&\,V \left(x^{j}\,-\,i \hbar \,\theta^{jl}\,\frac{\partial}{\partial x^{l}}\right)\Psi(x, t)\,.
\eea

\noindent
Refs.\cite{Chaichian1,Gamboa1,Gamboa2,Girotti1} illustrate the solving of Eq.(\ref{II-16}) for some specific models.

\section{Born series in noncommutative quantum mechanics}
\label{sec:level3}

Let us return, for a while, to commutative quantum mechanics and consider a system whose dynamics is described by the self-adjoint Hamiltonian operator

\be
\label{III-1}
H\,=\,H_0\,+\,V\left(X\right)\,,
\ee

\noindent
where $H_0 \equiv K_lK_l/2M$ will be referred to as the free Hamiltonian. Notice that $H = H^{\dagger}$ enforces $V = V^{\dagger}$ since the kinetic energy part $H_0$ is, by definition, self-adjoint. By construction $H_0$ does not possess bound states and its continuum energy spectrum is characterized by $E > 0$. By assumption, the same applies for the continuum spectrum of $H$ although this operator may also possess bound states. Furthermore, we shall set from now on $\hbar = 1$.

For the quantum system under consideration all observables can be obtained from the operator $T(W)$ defined by the integral equation

\be
\label{III-2}
T(W)\,=\,V\,+\,V\,G_0^{(+)}(W)\,T(W)\,,
\ee

\noindent
where $G_0^{(+)}(W)\,=\,\left[W\,-\,H_0\,+\,i\ve\right]^{- 1}$ is the free Green function for outgoing boundary conditions. By iterating the right hand side of Eq.(\ref{III-2}) one obtains $T$ as a series,

\be
\label{III-4}
T(W)\,=\,V\,+\,V\,G_0^{(+)}(W)\,V\,+\, V\,G_0^{(+)}(W)\,V\,G_0^{(+)}(W)\,V \,+\,\cdots\,,
\ee

\noindent
known as the Born series.

The problem of determining the necessary and sufficient conditions for the Born series to converge was solved by Weinberg\cite{Weinberg} long ago. He considers the eigenvalue problem

\be
\label{III-5}
G_0^{(+)}(W)\,V\,|\psi_{\nu}(W)\rangle\,=\,\eta_{\nu}(W)\,|\psi_{\nu}(W)\rangle\,.
\ee

\noindent
Since the operator $G_0^{(+)}(W)\,V$ is not Hermitean, the eigenvalues $\eta(W)$ may be complex. As for the eigenstates, $|\psi_{\nu}(W)\rangle$, they are assumed to be of finite norm. $W$ is kept negative or complex and is allowed to approach the positive real axis from above. From Eqs.(\ref{III-4}) and (\ref{III-5}) one obtains

\be
\label{III-6}
T(W)\,|\psi_{\nu}(W)\rangle\,=\,\left[\sum_{n = 0}^{\infty} \eta^n_{\nu}(W) \right]\,V\,|\psi_{\nu}(W)\rangle\,.
\ee

\noindent
It was demonstrated by Weinberg\cite{Weinberg} that

\be
\label{III-7}
|\eta_{\nu}(W)|\,<\,1\,,\qquad \forall \,\nu\,,
\ee

\noindent
is a necessary and sufficient condition for the Born series to converge.

We now want to solve the analogous problem for noncommutative quantum mechanics, the essential difference from the above being that instead of $V = V(X)$ we have $V = V(X^l + \theta^{lj}\,K_j)$. As point of departure, we start by invoking (\ref{III-5}) to cast Eq.(\ref{III-7}) as

\bea
\label{III-8}
&&\frac{|\langle {\vec k}\,|G_0^{(+)}(W)\,V\,|\psi_{\nu}(W)\rangle|}{|\langle {\vec k}\,|\psi_{\nu}(W)\rangle|}\,=\,\frac{1}{|W - \frac{{\vec k}^2}{2 M}\,+\,i \ve|}\,\frac{|\langle {\vec k}\,|\,V\,|\psi_{\nu}(W)\rangle|}{|\langle {\vec k}\,|\psi_{\nu}(W)\rangle|}\nonumber\\
&&\,=\,\frac{1}{|W - \frac{{\vec k}^2}{2 M}\,+\,i \ve|}\,\frac{1}{|\langle {\vec k}\,|\psi_{\nu}(W)\rangle|}\bigg|\int d^Nk^{\prime}\,\langle {\vec k}\,|\,V\,|{\vec k}^{\,\prime}\rangle \langle{\vec k}^{\,\prime}|\psi_{\nu}(W)\rangle\bigg|\nonumber\\
&&<\,1\,,\qquad \forall \,\nu\,.
\eea

\noindent
Let us concentrate on the linear momentum integral in the right hand side of Eq.(\ref{III-8}). Since

\be
\label{III-9}
\bigg|\int d^Nk^{\prime}\,\langle {\vec k}\,|\,V\,|{\vec k}^{\,\prime}\rangle \langle{\vec k}^{\,\prime}|\psi_{\nu}(W)\rangle\bigg| \,\leq\,\int d^Nk^{\prime}\,\big|\langle {\vec k}\,|\,V\,|{\vec k}^{\,\prime}\rangle \langle{\vec k}^{\,\prime}|\psi_{\nu}(W)\rangle\big|\,,
\ee

\noindent
one concludes that

\bea
\label{III-10}
\frac{1}{|W - \frac{{\vec k}^2}{2 M}\,+\,i \ve|}\,\frac{1}{|\langle {\vec k}\,|\psi_{\nu}(W)\rangle|}\,\int d^Nk^{\prime}\,\big|\langle {\vec k}\,|\,V\,|{\vec k}^{\,\prime}\rangle \big| \,\big|\, \langle{\vec k}^{\,\prime}|\psi_{\nu}(W)\rangle\big|\,<
\,1\,\qquad \forall \,\nu
\eea

\noindent
is a sufficient although not necessary condition for the convergence of the Born series. In other words, (\ref{III-10}) selects a subset of potentials for which the Born series certainly converge.

To proceed further on we shall be needing $\big|\langle {\vec k}\,|\,V\,|{\vec k}^{\,\prime}\rangle\big|$. Then, we start by looking for

\bea
\label{III-11}
&&\langle {\vec k}\,|\,V(X^l + \theta^{lj}\,K_j)\,|{\vec k}^{\,\prime}\rangle\,=\,\int d^Nx\,\phi_{\vk}^{\star}(\vx)\,V\left(x^l\,-\,i\,\,\theta^{lj}\,\frac{\pa}{\pa x^j}\right)\,\phi_{{\vk}^{\,\prime}}(\vx)\nonumber\\
&&=\,\int d^Nx\,\phi_{\vk}^{\star}(\vx)\,\left[V(\vx) \star \phi_{{\vk}^{\,\prime}}(\vx)\right]\,=\,\int d^Nx\,\phi_{\vk}^{\star}(\vx) \star V(\vx) \star \phi_{{\vk}^{\,\prime}}(\vx)\nonumber\\
&&=\,\int d^Nx\, V(\vx)\,\left[\phi_{{\vk}^{\,\prime}}(\vx) \star \phi_{\vk}^{\star}(\vx) \right]\,,
\eea

\noindent
where
\be
\label{III-12}
\phi_{{\vk}}(\vx) = \frac{1}{(2 \pi )^{\frac{N}{2}}}\,e^{i\,k_j x^j}\,,
\ee

\noindent
is the eigenfunction of the linear momentum ${\vec K}$, corresponding to the eigenvalue ${\vk}$. From Eq.(\ref{II-17}) one, then, finds

\bea
\label{III-13}
\phi_{{\vk}^{\,\prime}}(\vx) \star \phi_{\vk}^{\star}(\vx)\,=\,\phi_{{\vk}^{\,\prime}}(\vx) \,\left[\exp \left(- i
 \overleftarrow{\frac{\pa}{\pa
x^l}} \,\theta^{lj}\, \overrightarrow{\frac{\pa}{\pa
x^{j}}}\right)\right]\,\phi_{\vk}^{\star}(\vx)\,=\,e^{-\,i\,{\vec k}^{\,\prime}\,\wedge\,\vk}\,\phi_{{\vk}^{\,\prime}}(\vx)\, \phi_{\vk}^{\star}(\vx)\,,
\eea

\noindent
where

\be
\label{III-131}
{\vec k}^{\,\prime}\,\wedge\,\vk\,\equiv\,k_l^{\prime}\,\theta^{l j}\,k_j\,.
\ee

\noindent
Clearly, Eqs.(\ref{III-13}) and (\ref{III-11}) amount to

\be
\label{III-14}
\langle {\vec k}\,|\,V(X^l + \theta^{lj}\,K_j)\,|{\vec k}^{\,\prime}\rangle\,=\,e^{-\,i\,{\vec k}^{\,\prime}\,\wedge\,\vk}\,\langle {\vec k}\,|\,V(X^l)\,|{\vec k}^{\,\prime}\rangle\,
\ee

\noindent
and, as consequence,

\be
\label{III-15}
\big|\langle {\vec k}\,|\,V(X^l + \theta^{lj}\,K_j)\,|{\vec k}^{\,\prime}\rangle\big|\,=\, \big|\langle {\vec k}\,|\,V(X^l)\,|{\vec k}^{\,\prime}\rangle \big|\,.
\ee

This result connects the commutative with the noncommutative regimes. Therefore, if $V(X)$ verifies Eq.(\ref{III-10}) so does $V(X^l + \theta^{lj}\,K_j)$ or, what amounts to the samething, for the restricted subclass of potentials verifying Eq.(\ref{III-10}) the convergence of the Born series holds for both, the commutative and the noncommutative versions of the model.

\section{Unitarity in noncommutative quantum mechanics}
\label{sec:level4}

The scattering amplitude $f( {\vk}^{\,\prime}\,,\,\vk)$ is given in terms of the $T$-matrix by

\be
\label{III-16}
f( {\vk}^{\,\prime}\,,\,\vk)\,\equiv\,-\,4 \pi^2 M\,T( {\vk}^{\,\prime}\,,\,\vk)\,,
\ee

\noindent
where $T( {\vk}^{\,\prime}\,,\,\vk)$ is short for $\langle {\vk}^{\,\prime}|T|\vk \rangle$. Unitarity demands that

\be
\label{III-17}
\Im\,f( {\vk}\,,\,\vk)\,=\,\frac{k}{4 \pi}\,\int\,d\Omega_{\vk^{\,'}}\,\big|f( {\vk}^{\,'}\,,\,\vk)\big|^2\,,
\ee

\noindent
where $k = |\vk|$ and $d\Omega_{\vk^{\,'}}$ is the element of solid angle centered around $\vk^{\,'}$.

Our purpose here is to check (\ref{III-17}) by taking advantage of the Born series representation for $T$. It will be assumed that the potential $V$ contains a dimensionless coupling constant ($g$) that enables one to write $V = g\, U$. Then, the Born series in Eq.(\ref{III-4}) becomes a power series in $g$. Correspondingly, Eq.(\ref{III-17}) translates into

\be
\label{III-19}
\frac{4\,\pi}{k}\,\Im\,f^{(n)} ({\vk}\,,\,\vk)\,=\,\int\,d\Omega_{\vk^{\,'}}\,\sum_{i = 1}^n\,f^{(i)^{\star}}( {\vk}^{\,'}\,,\,\vk)\,f^{(n\,-\,i)}( {\vk}^{\,'}\,,\,\vk)\,,
\ee

\noindent
where $n$ is a positive integer,

\be
\label{III-191}
f^{(n)}( {\vk}^{\,'}\,,\,\vk)\,=\,-\,4 \pi^2 M\,T^{(n)}(\vk\,' , \vk)\,,
\ee

\noindent
and

\bea
\label{III-192}
T^{(n)}(\vk , \vk\,')\,=\,\langle \vk\, \big|\overbrace{V G_0^{(+)}(E)V\cdots V G_0^{(+)}(E)V}^{n\, \mbox{factors}\, V; (n - 1)\, \mbox{factors}\,G_0^{(+)}(E)}  \big|\vk\,'\rangle\,.
\eea

Let us first analyze the contributions to the scattering amplitude for $n = 1$. Since the right hand side in (\ref{III-19}) does not contain terms of order $g^1$ no term of such order should arise in $\Im\,f^{(1)}( {\vk\,,\,\vk})$. We know that this is the case in the commutative version of the theory, since the hermiticity of $V$ secures $\Im \,\langle {\vec k}\,|\,V(X^l)\,|{\vec k}\rangle\,=\,0$. As for the noncommutative case, we observe that for $\vk^{\,\prime}\,=\,\vk$ (forward direction) the exponent in the right hand side of (\ref{III-14}) vanishes and, therefore, $\Im \,\langle {\vec k}\,|\,V(X^l + \theta^{lj}\,K_j)\,|{\vec k}\rangle\,=\,\Im \,\langle {\vec k}\,|\,V(X^l)\,|{\vec k}\rangle\,=\,0$, as required.

To verify Eq.(\ref{III-19}) for arbitrary $n$ we start by claiming that

\bea
\label{III-20}
&&\Im\,\int\,d^Nk^{\prime}\,\frac{T^{(m)^{\star}}(\vk\,' , \vk)\,T^{(p)}(\vk\,' , \vk)}{\frac{k^2}{2M}\,-\,\frac{k^{'\,2}}{2M}\,+\,i\ve}\,=\,
\Im\,\int\,d^Nk^{\prime}\,\frac{T^{(m + 1)^{\star}}(\vk\,' , \vk)\,T^{(p - \,1)}(\vk\,' , \vk)}{\frac{k^2}{2M}\,-\,\frac{k^{'\,2}}{2M}\,+\,i\ve}\nonumber\\
&&-\,M\,k\,\pi\,\int\,d\Omega_{\vk\,'}\,\left[ T^{(m)^{\star}}(\vk\,' , \vk)\,T^{(p)}(\vk\,' , \vk)\,+\,T^{(p)^{\star}}(\vk\,' , \vk)\,T^{(m)}(\vk\,' , \vk)\right]\,,
\eea

\noindent
whose proof is straightforward but will be omitted for reasons of space. Then, consider

\bea
\label{III-21}
&&\Im T^{(n)}(\vk , \vk)\,=\,\Im \int\,d^Nk^{\prime}\,\frac{T^{(1)^{\star}}(\vk\,' , \vk)\,T^{(n - 1)}(\vk\,' , \vk)}{\frac{k^2}{2M}\,-\,\frac{k^{'\,2}}{2M}\,+\,i\ve}\nonumber\\
&&=\,\Im\,\int\,d^Nk^{\prime}\,\frac{T^{(2)^{\star}}(\vk\,' , \vk)\,T^{(n - \,2)}(\vk\,' , \vk)}{\frac{k^2}{2M}\,-\,\frac{k^{'\,2}}{2M}\,+\,i\ve}\nonumber\\
&&-\,M\,k\,\pi\,\int\,d\Omega_{\vk\,'}\,\left[ T^{(1)^{\star}}(\vk\,' , \vk)\,T^{(n - \,1)}(\vk\,' , \vk)\,+\,T^{(n - \,1)^{\star}}(\vk\,' , \vk)\,T^{(1)}(\vk\,' , \vk)\right]\,,
\eea

\noindent
where in going from the second to the third term of the equality we have used (\ref{III-20}) for $ m = 1$ and $p = n - 1$. It is not difficult to see that by applying this procedure $ (n - 2)$ times one ends up with

\bea
\label{III-22}
&&\Im T^{(n)}(\vk , \vk)\,=\,\Im T^{(n)^{\star}}(\vk , \vk)\nonumber\\
&&- 2 M k \pi\,\int\,d\Omega_{\vk\,'}\,\left[ T^{(1)^{\star}}(\vk\,' , \vk) T^{(n - \,1)}(\vk\,' , \vk) + \cdots + T^{(n - \,1)^{\star}}(\vk\,' , \vk) T^{(1)}(\vk\,' , \vk)\right]\,,
\eea

\noindent
which, after recalling that $\Im T^{(n)^{\star}}(\vk , \vk)\,=\,-\,\Im T^{(n)}(\vk , \vk)$,  goes into

\bea
\label{III-23}
\Im T^{(n)}(\vk , \vk)\,=\,-\,M \,k \,\pi\,\int\,d\Omega_{\vk\,'}\,\sum_{i = 1}^n\,T^{(i)^{\star}}( {\vk}^{\,'}\,,\,\vk)\,T^{(n\,-\,i)}( {\vk}^{\,'}\,,\,\vk)\,.
\eea

\noindent
This last equation reproduces Eq.(\ref{III-19}) in terms of $T$-matrix elements and, hence, concludes the purported proof of unitarity. It applies equally well for the commutative and the noncommutative cases.

\section{Conclusions}
\label{sec:level5}

This work was dedicated to demonstrate that some of the essential ingredients of the commutative version of quantum mechanics remain valid in the noncommutative counterpart.

Our first concern was about the existence of the Born series, since it provides the most powerful tool for calculating the $T$-matrix. We made explicit the condition to be fulfilled by the potential to that end.

The fact that the non-commutativity does not destroy the Born series greatly facilitated the proof of unitarity, which is an essential requirement for a quantum theory to make sense.

We believe that the results presented in this paper contribute to support noncommutative quantum mechanics as a sound quantum theory.\\

\noindent
{\bf ACKNOWLEDGEMENTS}\\

Both of us acknowledge partial support from Conselho Nacional de Desenvolvimento Cient\'{\i}fico e Tecnol\'ogico (CNPq), Brazil.


\begin{thebibliography}{99}
\bibitem {footnote1} Observe that the dimension of $\theta^{lk}$ in the cgs system is $d[\theta] =$ g$^{-1}$ s.
\bibitem {footnote2} The functional formulation of the quantum dynamics of noncommutative systems is out of the scope  of this paper.
\end{thebibliography}
\end{document}